\documentclass[submission, Phys]{SciPost}

\usepackage{amsmath}
\usepackage{amssymb}
\usepackage{amsbsy}
\usepackage{graphicx}
\usepackage{color}
\usepackage{lmodern}

\usepackage{enumerate}
\usepackage{mathtools}
\usepackage{ulem}

\def\be{\begin{equation}}
\def\ee{\end{equation}}
\def\bfi{\begin{figure}}      
\def\efi{\end{figure}}
\def\bea{\begin{eqnarray}}
\def\eea{\end{eqnarray}}

\begin{document}

\begin{center}{\Large \textbf{Estimating generation time of SARS-CoV-2  variants from the daily incidence rate}}\end{center}

\begin{center}
  E. Lippiello\textsuperscript{1},
  G. Petrillo\textsuperscript{2},
  L. de Arcangelis\textsuperscript{1},
\end{center}

\begin{center}
  {\bf 1}Department of Mathematics and Physics, University of Campania ``Luigi Vanvitelli'', 81100, Caserta, Italy
\\
  {\bf 2} The Institute of Statistical Mathematics, Research Organization of Information and Systems, Tokyo, Japan 
\\

* eugenio.lippiello@unicampania.it
\end{center}

\begin{center}
\today
\end{center}


\section*{Abstract}
         {\bf The identification of the transmission parameters of a virus is fundamental to identify the optimal public health strategy. These parameters can present significant changes over time caused by 
           genetic mutations or viral recombination, making their continuous  monitoring fundamental. Here we present a method, suitable for this task, which uses as unique information the daily number of reported cases. The method is based on a time since infection model where transmission parameters are obtained by means of an efficient maximization procedure of the likelihood. Applying the method to SARS-CoV-2  data in Italy we find an average generation time $\overline{z}=3.2 \pm 0.8$ days, during the temporal window when the majority of infections can be attributed to the Omicron variants. At the same time we find a significantly larger value $\overline{z}=6.2\pm 1.1$ days,  in the temporal window when spreading was dominated by the Delta variant. We are also able to show that the presence of the Omicron variant, characterized by a shorter $\overline z$, was already detectable in the first weeks of December 2021, in full agreement with results provided by sequences of SARS-CoV-2 genomes reported in national databases. Our results therefore indicate that the novel approach can indicate the existence of virus variants resulting particularly useful in situations when information about genomic sequencing is not yet available.}

\noindent\rule{\textwidth}{1pt}
\tableofcontents\thispagestyle{fancy}
\noindent\rule{\textwidth}{1pt}
\vspace{10pt}

\section{Introduction}

SARS-CoV-2, as other viruses, are continuously evolving because of genetic mutations or viral recombination. These changes can strongly affect transmission parameters \cite{AndMay02}  inducing important differences in the virus spreading. 
In particular a reduction of the generation time $z$, i.e. the time difference between the dates of infection of successive cases in a transmission chain, leads to an increased epidemic growth rate, even for unaltered  reproduction number $R_0$. Furthermore, an accurate estimate of the mean value of the generation time $\overline{z}$ is fundamental to establish the optimal duration of the quarantine period.  

Elegant methods based on log-likelihood maximization have been recently developed \cite{Ganyetal20,Ferretal20,Ferretal20b,LPdA22} to obtain the average value 
$\overline{z}$ of the generation time. However, very often $\overline{z}$ is identified with $\overline{s}$, defined as the mean value of the serial interval  \cite{Ferretal20b}, which is the difference in timing of symptom onset in a pair of a primary and its secondary case.
The measurement of $\overline{s}$, indeed, differently from the measurement of $\overline{z}$, can be directly obtained from the reconstruction of the contact network. This information, combined with the results provided  by genomic sequencing, provides an estimate of the mean value of serial intervals of each specific  variant \cite{Bacetal22}. Nevertheless, it is important to remark \cite{CD15,Alietal20,Parketal21,Manietal22}  that the value of $\overline s$, obtained from contact tracing can be significantly different than the ``intrinsic'' value of $\overline z$. This occurs, in particular,  when the structure of the contact network fastly changes in time, as for instance in presence of non-pharmaceutical interventions.  
Here ``intrinsic'' refers to the quantity measured in the ideal case of a fully susceptible, homogeneously mixed population \cite{CD15}, and therefore independent of the specific conditions of the epidemiological setting from which it is inferred.

In this study we will show that the intrinsic value of $\overline{z}$  can be obtained by means of a completely data driven procedure. The main observation is that, if the value of  $\overline{z}$
affects the future evolution of the number of infected cases, its value could be potentially extracted from the previous evolution of the virus spreading. More precisely we use the method recently developed by us \cite{LPdA22} to extract $\overline{z}$ directly from  the daily series of incidence rate ${I(t)}$, i.e. the number of   infected individuals at the calendar time $t$. The method is based on the non-trivial dependence on $\overline{z}$ of the Log-Likelihood function $LL$, which measures the overlap between the measured  ${I(t)}$ and the expected one, according to a time since infection model \cite{KMKW27,GrFr08}. We show that when two variants with differences in $\overline{z}$ are simultaneously present in the sample, $LL(\overline{z})$ presents two distinct peaks in correspondence to the mean value of $z$ of the two dominant variants. Furthermore the ratio between the two peak heights also provide information about the relative incidence of the two variants in virus spreading.

We perform this study using the incidence rate ${I(t)}$ for SARS-CoV-2  in Italy where three Variants of Concern (VOC) have dominated in three different temporal windows, as clearly highlighted by Fig.\ref{fig1} extracted from the Bulletin n.18 of March 25th 2022 of Istituto superiore di Sanità (https://www.epicentro.iss.it/coronavirus/SARS-CoV-2-monitoraggio-varianti-rapporti-periodici). More precisely more than the $90\%$ of total infections in Italy can be attributed to the Alpha variant in the temporal window from 2021-03-08 to 2021-05-17, to the Delta variant in the temporal window from 2021-07-12  to 2021-11-29, and to the Omicron variant in the temporal window subsequent to 2022-01-10.

\begin{figure*}
  \includegraphics[width=15cm,height=8cm]{fig1_variant.png}
  \caption{Major variants identified by sequencing provided by the I-Co-Gen platform software, 
    by weekly sampling (December 28, 2020 - March 21, 2022). Data after the dotted line must be considered not consolidated. The figure is extracted from ``Prevalenza e distribuzione delle varianti di SARS-CoV-2 di interesse 
per la sanità pubblica in Italia'', Rapporto n. 18 del 25 marzo 2022, https://www.epicentro.iss.it/coronavirus/SARS-CoV-2-monitoraggio-varianti-rapporti-periodici. }
  \label{fig1}
\end{figure*}

Several studies \cite{Braetal21,Bacetal22,Songetal22,aHB22} have measured, in different geographic regions, a value $\overline s$ of the Omicron variant significantly shorter than the value measured for previous variants Alpha and Delta. Because of this observation,  many countries have applied a reduction of the duration of the quarantine period (https://www.ecdc.europa.eu/en/news-events/ecdc-updates-guidance-regarding-quarantine-andisolation-considering-spread-of-omicr). This is in agreement with the evaluation of the intrinsic value of $\overline z$ using nucleotide sequences of SARS-CoV-2 viruses sampled in Denmark \cite{IPN22}, leading to a value of $\overline z$ for the Omicron variant about $0.5-0.6$ times smaller than the one measured for the Delta variant. Conversely, a study of infections among household members in Reggio Emilia (Italy) lead to an intrinsic value of $\overline z$  which is about $6$ days, with no significant difference among the three variants, Alpha, Delta e Omicron \cite{Manietal22}.
Our data appear more consistent with the result of \cite{IPN22} since we find a value $\overline z=3.2 \pm 0.8$ days to the Omicron variant respect to the value $\overline z=6.5 \pm 1$ days associated to the Delta one.

\section{The method}  

In this section we overview the method considered in this study, details can be found in \cite{LPdA22}.

The starting point is the renewal equation \cite{KMKW27,Fra07,GrFr08} providing the expected value of daily infected people on the $m$-th day, $E[I(m)]$, in terms of the past daily incidence 
\be
E[I(m)]=\sum_{j=0}^{m-1}R_c(j) w(m-j)I(j) +\mu(m),
\label{rate2}
\ee
where $R_c(m)$ is the case reproduction number, representing the total number of infections induced on average by an individual infected on the $m$-th day, $w(j)$ is the distribution of generation times, representing the percentage of infections induced at a time distance $j$ from the infection and, finally,
$\mu(m)$ is the daily number of imported cases during the $m$-th day, i.e. infectors coming from outside the considered region. We assume that $w(j)$ is a Gamma distribution, $w(j)=\left(\tau^{-a}/\Gamma(a) j^{a-1}\right) \exp(-j/\tau)$, which depends on two parameters, $a\ge 1$ and $\tau>0$, and where $\Gamma(a)$ is the Gamma function. The  Gamma distribution is fully characterized by its average value $\overline z$ and by its standard  deviation $\sigma$, which are both functions  of $a$ and $\tau$, $\overline z=a \tau$ and  $\sigma=\sqrt{a} \tau$. 
Similar results are found for a  Weibull or a log-normal distribution \cite{LPdA22}.

An analytical expression for the log-likelihood $LL$  of the time series $\{I(m)\}_{m=1,...,N}$, for  assigned sequences $\{R_c(m)\}_{m=1,..,N}$, $\{\mu(m)\}_{m=1,..,N}$, and for given values of  $\overline z$ and $\sigma$ has been obtained \cite{LPdA22} under the hypothesis that the number of individuals infected on the $m$-th day  is Poisson distributed. For fixed $\overline z$ and $\sigma$, the best series  $\{R_c(m)\}_{m=1,..,N}$ and $\{\mu(m)\}_{m=1,..,N}$  which maximize $LL$ are finally obtained by generalizing the  Markov-chain-Monte-Carlo method introduced to find the optimal parameters in epidemic models for seismic occurrence \cite{BLGdA11,LGdAMG14}. 

In the following we define $LL^{best}(\overline z,\sigma)$, the value of $LL$ in correspondence to the best series  $\{R_c(m)\}_{m=1,..,N}$ and $\{\mu(m)\}_{m=1,..,N}$ and we explore its dependence 
on the parameters $\overline z$ and $\sigma$.

\section{Results}

\begin{figure}
  \includegraphics[width=15cm,height=7cm]{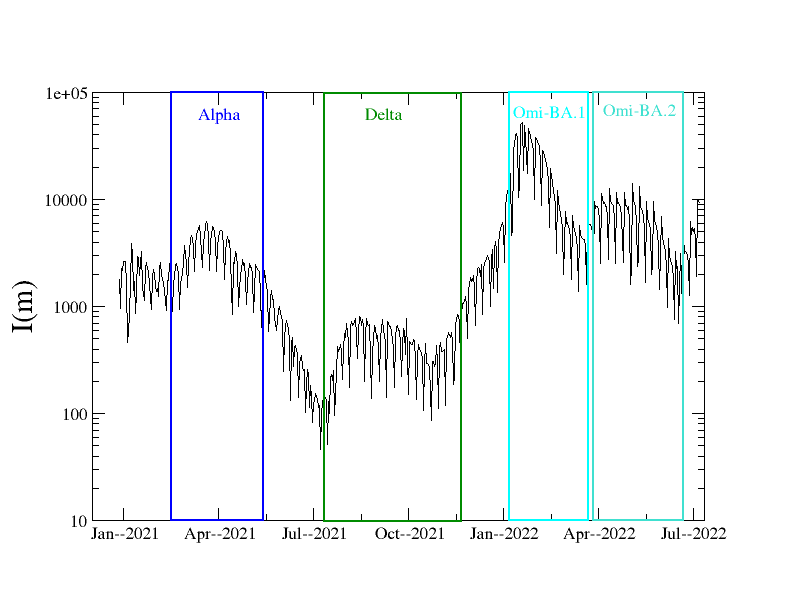}
  \caption{The daily incidence $I(m)$ of SARS-CoV-2 in Lombardy from January 2021 up to June 2022. Colored boxes identify the four different temporal windows Alpha, Delta, Omicron-BA.1 and Omicron-BA.2 obtained from Fig.\ref{fig1}.}
    \label{fig2}
\end{figure}

We consider data provided by the Department of Protezione Civile in Italy. More precisely we mainly consider data for the region Lombardy where the first outbreak of SARS-CoV-2 has been documented in Europe and which is characterized by a widespread diffusion of the disease since March 2020. In Fig.\ref{fig2} we plot the daily incidence from January 2021. In the figure we highlight the three main temporal windows which, according to the results of Fig.\ref{fig1}, are mostly characterized by the spreading of a specific variant. We have also identified two sub-windows where the spreading is mainly controlled by two different  lineages of Omicron BA.1 and BA.2, respectively.   
It is evident that each temporal window corresponds to a different wave of Covid spreading with a distinct peak in $I(m)$.

\begin{figure*}
  \includegraphics[width=15cm,height=12cm]{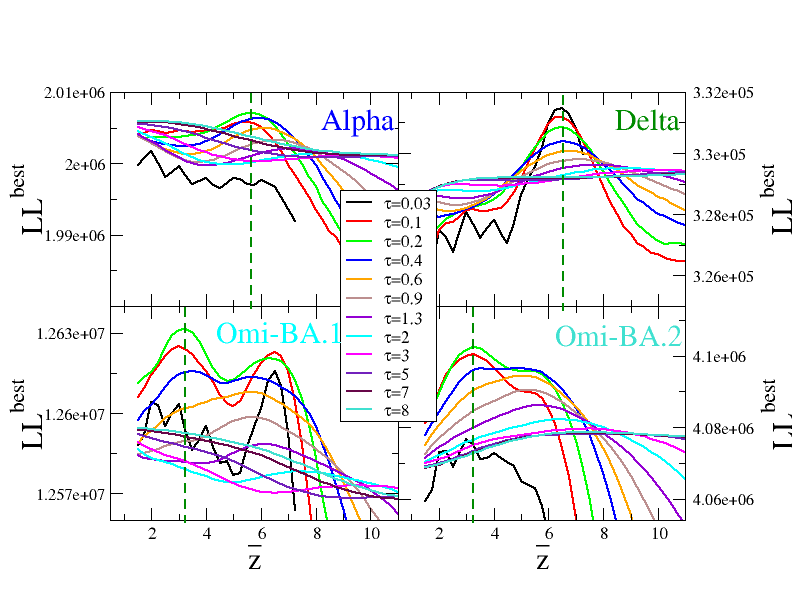}
  \caption{ The log-likelihood $LL^{best}(\overline z,\sigma)$, evaluated for the temporal profile of $R_c(m)$ which maximizes the likelihood for the daily incidence of SARS-CoV-2 in Lombardy, is plotted as a function of $\overline z=a \tau$. The four different panels correspond to the four temporal windows highlighted in Fig.\ref{fig2}. 
    Different curves, in each panel, correspond to different values of $\tau$,  which implies a different $\sigma=a \sqrt{\tau}$. The dashed green vertical line identifies the value of $\overline z$ which provides the maximum value of the log-likelihood, in each panel.} 
    \label{fig3}
\end{figure*}

We separately apply the procedure outlined in the previous section, restricting to data within each of the $4$ temporal windows, which are classified as Alpha, Delta, Omicron-BA.1 and Omicron-BA.2.
The values of  $LL^{best}(\overline z,\sigma)$, for different choices of $\overline z$ and $\sigma$,  for each of the four temporal windows are plotted in a separate panel of Fig.\ref{fig3}. More precisely, we plot $LL^{best}(\overline z,\sigma)$ versus $\overline z$ and different values of the parameter $\tau$ of the Gamma distribution. Results clearly show that during the Alpha window,  $LL^{best}(\overline z,\sigma)$ presents a clear maximum for $\overline z=5.7$ days when $\tau=0.2$ days, leading to an estimate $\overline z=5.7$ days and $\sigma=1.1$ days, consistently with previous findings both in terms of serial interval and intrinsic generation time. During the Delta period the peak is even more pronounced at $\overline z=6.5$ days for $\tau=0.03$ days, consistently with previous results. Interestingly, during the Omicron-BA.1 window the maximum of $LL^{best}$ at $\overline z=6.5$, observed during the Delta window,  is still present but is subleading, and the most relevant peak is present  at a significantly smaller values   $\overline z=3.2$ days for $\tau=0.2$ days. During the Omicron-BA.2 window the peak at $\overline z=3.2$ days is the only relevant one presented  by  $LL^{best}$.
Results of Fig.\ref{fig3} clearly show a significant reduction of the generation time of the Omicron variants, with an estimated value $\overline z =3.2$ days with $\sigma=0.8$ days, which is roughly half of the value estimated during the Delta period, in agreement with results of serial intervals and of ref.\cite{IPN22}. On the other hand, we do not find significant differences for the value of $\overline z$ between the  two Omicron lineages BA.1 and BA.2.

We remark that in the presence of two peaks of $LL^{best}$, if the range of parameters is not completely explored, it may occur that automatic procedures for log-likelihood maximization, based on Monte Carlo Markov Chains, could remain trapped in a relative maximum without reaching the global one. The result of ref.\cite{Manietal22} could be affected by this problem identifying as  best model parameters the ones related to the Delta peak instead of those associated to the  Omicron one.   

\begin{figure*}
  \includegraphics[width=15cm,height=12cm]{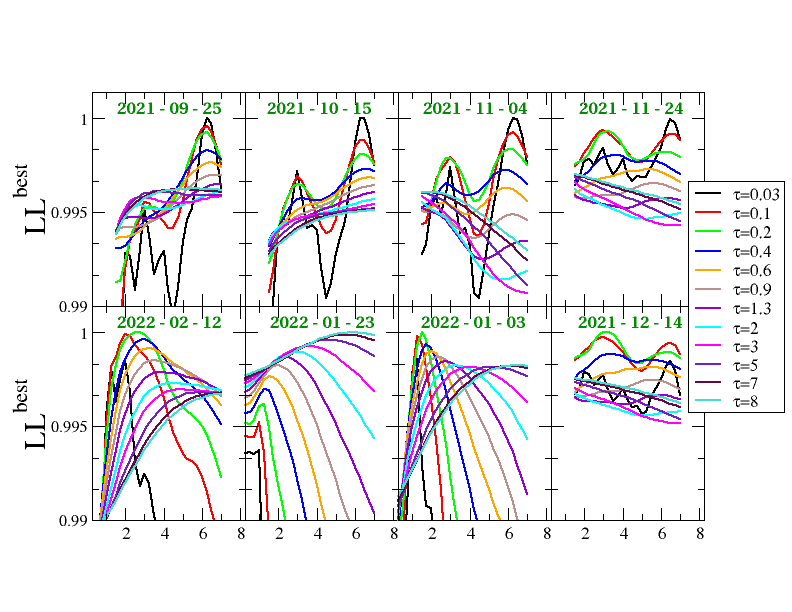}
  \caption{ The log-likelihood $LL^{best}(\overline z,\sigma)$ is plotted as a function of $\overline z=a \tau$. Different panels correspond to different temporal windows of $60$ days, with the initial day of each window reported on the top of each panel. The value of $LL^{best}(\overline z,\sigma)$ has ben divided by its maximum value $LL^{best}(\overline z^{max},\sigma^{max})$ in each temporal window, so that the maximum value of $LL^{best}$ is normalized to $1$ in each panel.
Panels are organized in such a way that the initial time of temporal windows increases moving from left to right in upper panels and then continuous to increase moving in the lower panel from right to left. Different curves, in each panel, correspond to different values of $\tau$,  which implies a different $\sigma=a \sqrt{\tau}$.}
      \label{fig4}
\end{figure*}


In Fig.\ref{fig4} we present the behavior of $LL^{best}(\overline z,\sigma)$ as function of $\overline z$ within temporal windows of a fixed duration of $60$ days, with different starting days, ranging  from the first one which is fully contained within the Delta window up to the last one which is fully inside the Omicron one.
Results show that in the temporal window starting on 2021-09-25 (first upper panel) only the peak at $\overline z \simeq 6.5$ is visible in $LL^{best}$. By shifting forward the starting time and considering a time window starting on 2021-10-15 (second upper panel) a subleading peak at $\overline z \simeq 3$ appears. This second peak in $LL^{best}$ therefore signals the presence of a new variant with a different $\overline z$ in the first weeks of December 2021. This is fully consistent with the results of Fig.\ref{fig1} indicating that the percentage of infections caused by the Omicron variant starts to be significant in the first weeks of December 2021. Moreover, consistently with Fig.\ref{fig1}, Fig.\ref{fig4} shows that by further shifting forward the starting day (upper panels form left to right) the peak at $\overline z \simeq 3$ becomes increasingly more relevant until it turns on the dominant one in the temporal window starting on 2021-12-14 (fourth lower panel). This is again consistent with the results of Fig.\ref{fig1} indicating that the Omicron is the most relevant variant after the mid-December 2021. Keeping on shifting forward the starting time (lower panels from right to left) the peak at $\overline z \simeq 3$ becomes more visible remaining the only relevant one in $LL^{best}(\overline z,\sigma)$ in the time window starting on 2022-02-12. We remark that no clear indication can be extracted from $LL^{best}(\overline z,\sigma)$ in the temporal window starting on 2022-01-23. In this case indeed a clear peak is not visible and the largest value of $LL^{best}$ is obtained for the largest considered value of $\tau=8$ days, indicating that the standard deviation can be as large as $10$ days and therefore does not allow us to obtain any information on $\overline z$. We have no clear justification for the very peculiar behavior of $LL^{best}$ in this temporal window, which corresponds to the period when $I(m)$ is in a fast  decreasing phase. It could be possible that new infections within this time window are too few to extrapolate transmission parameters from $I(m)$.

\begin{figure*}
  \includegraphics[width=15cm,height=12cm]{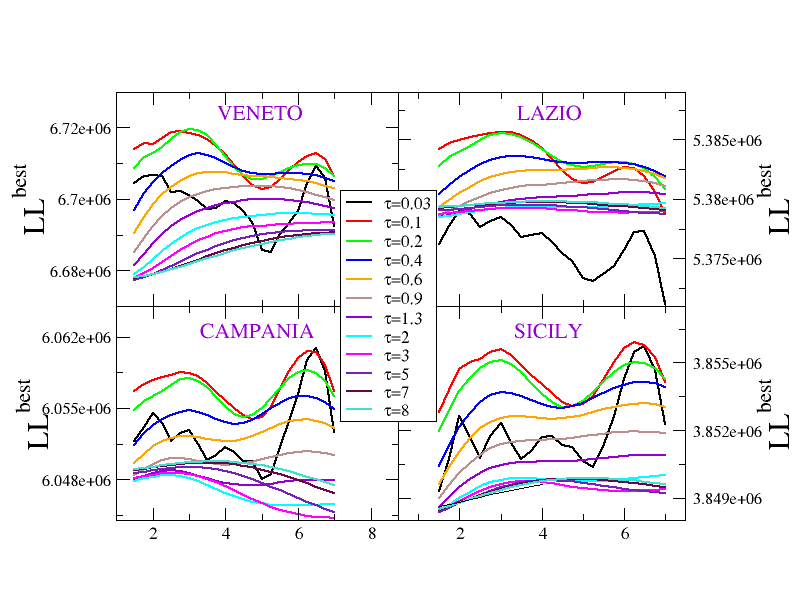}
  \caption{The log-likelihood $LL^{best}(\overline z,\sigma)$ is plotted as a function of $\overline z=a \tau$ during the Omicron-BA.1 temporal window, for four Italian regions: Veneto (upper left panel), Lazio (upper right panel), Campania (lower left panel) and Sicily (lower right panel).
     Different curves, in each panel, correspond to different values of $\tau$,  which implies a different $\sigma=a \sqrt{\tau}$.}
      \label{fig6}
\end{figure*}

In Fig.\ref{fig6} we consider the behavior of $LL^{best}(\overline z,\sigma)$ in different Italian regions during the Omicron-BA1 window. Results suggest the simultaneous presence of the two variants Delta and Omicron in all the considered regions. Indeed, in all regions the two peaks at $\overline z \simeq 3$ and 
$\overline z \simeq 6$ are clearly visible. However, the relevance of the two peaks is different among the different regions. Indeed, in some regions like Lazio,  the Omicron variant clearly appears as the dominant one during the Omicron-BA.1 window. Conversely, in Campania  the contagion appears still more controlled by the Delta variant whereas in Sicily the two variants appear to contribute in a similar way to SARS-CoV-2 diffusion. In Veneto, finally, one recovers a situation very similar to the one of Lombardy (Fig.\ref{fig3}) with a small predominance of the Omicron variant with respect to the Delta one.

\section{Conclusions}

We have considered an epidemic model based on a renewal equation (Eq.\ref{rate2}) which depends on the transmission parameters $R_c(m)$, representing the time dependent case reproduction number, and on the parameters $\overline{z}$ and $\sigma$, representing the mean value and the standard deviation, respectively, of the generation time distribution. We have used this model to describe the daily incidence rate of SARS-CoV-2 $I(m)$ in Italian regions during different temporal windows. More precisely, 
we have obtained the value of model parameters providing the best description of experimental data by using the log-likelihood maximization procedure introduced in ref. \cite{LPdA22}. In particular, we have separately considered data in four different temporal windows corresponding to periods  when the diffusion of SARS-CoV-2 was mostly controlled by one of the four variants (Alpha, Delta, Omicron-BA1 and Omicron-BA2). We have found that $\overline{z}$ during the Omicron windows was significantly smaller than, about one half of the value measured during Alpha and Delta windows, consistently with previous results about serial intervals \cite{Braetal21,Bacetal22,Songetal22,aHB22} and an estimate of  $\overline{z}$ in Denmark \cite{IPN22}.
By studying the behavior of the log-likelihood in different time windows, we find a clear indication of the presence of the Omicron variant in Italy since the first weeks of December 2021 with a diffusion becoming more and more relevant at later times. Our results are fully consistent with the relative diffusion of the different SARS-CoV-2 variants  identified by sequencing provided by the I-Co-Gen platform software over the Italian territory,  

Summarizing, our study shows that the adopted procedure can be very useful to identify, in about real time, changes in the transmission parameters of a virus that can be attributed to its mutations. We remark that this result can be obtained only from the daily number of infected individuals  without any further information  about the identification of the correct infector-infectee pair, ignoring the timing of symptom onsets as well as other details which are necessary to reconstruct the transmission chain in traditional approaches. More importantly, our approach does not need the support of laboratory analysis for genomic sequences, which is not always available. Accordingly, the procedure adopted in this manuscript could be particularly useful in the early stage of a new pandemic, or in the early stage  of a new mutation, when the genetic information on the virus is not yet complete and genomic classification is not yet available.

\section{Acknowledgments}
E. L. and G. P. acknowledge support from project PRIN201798CZLJ.  L. de A. acknowledges support from project PRIN2017WZFTZP. E.L. and L. de A. acknowledge support from VALERE project $E-PASSION$ of the University of Campania ``L. Vanvitelli''. G. P. acknowledges support from MEXT Project for Seismology TowArd Research innovation with Data of Earthquake (STAR-E Project), Grant Number: JPJ010217.


\begin{thebibliography}{10}
\providecommand{\url}[1]{\texttt{#1}}
\providecommand{\urlprefix}{URL }
\expandafter\ifx\csname urlstyle\endcsname\relax
  \providecommand{\doi}[1]{doi:\discretionary{}{}{}#1}\else
  \providecommand{\doi}{doi:\discretionary{}{}{}\begingroup
  \urlstyle{rm}\Url}\fi
\providecommand{\eprint}[2][]{\url{#2}}

\bibitem{AndMay02}
R.~M. Anderson and R.~M. May,
\newblock \emph{Infectious Diseases of Humans: Dynamics and Control},
\newblock Oxford Science Publications, Oxford (2002).

\bibitem{Ganyetal20}
T.~Ganyani, C.~Kremer, D.~Chen, A.~Torneri, C.~Faes, J.~Wallinga and N.~Hens,
\newblock \emph{Estimating the generation interval for coronavirus disease
  (Covid-19) based on symptom onset data, march 2020},
\newblock Eurosurveillance \textbf{25}(17), 2000257 (2020),
\newblock \doi{https://doi.org/10.2807/1560-7917.ES.2020.25.17.2000257}.

\bibitem{Ferretal20}
L.~Ferretti, C.~Wymant, M.~Kendall, L.~Zhao, A.~Nurtay, L.~Abeler-D{\"o}rner,
  M.~Parker, D.~Bonsall and C.~Fraser,
\newblock \emph{Quantifying SARS-CoV-2 transmission suggests epidemic control
  with digital contact tracing},
\newblock Science \textbf{368}(6491) (2020),
\newblock \doi{10.1126/science.abb6936},
\newblock
  \eprint{https://science.sciencemag.org/content/368/6491/eabb6936.full.pdf}.

\bibitem{Ferretal20b}
L.~Ferretti, A.~Ledda, C.~Wymant, L.~Zhao, V.~Ledda, L.~Abeler-D{\"o}rner,
  M.~Kendall, A.~Nurtay, H.-Y. Cheng, T.-C. Ng, H.-H. Lin, R.~Hinch
  \emph{et~al.},
\newblock \emph{The timing of Covid-19 transmission},
\newblock medRxiv  (2020),
\newblock \doi{10.1101/2020.09.04.20188516},
\newblock
  \eprint{https://www.medrxiv.org/content/early/2020/09/16/2020.09.04.20188516.full.pdf}.

\bibitem{LPdA22}
E.~Lippiello, G.~Petrillo and L.~de~Arcangelis,
\newblock \emph{Estimating the generation interval from the incidence rate, the
  optimal quarantine duration and the efficiency of fast switching periodic
  protocols for Covid-19},
\newblock Sci Rep \textbf{12}, 4623 (2022),
\newblock \doi{10.1038/s41598-022-08197-x}.

\bibitem{Bacetal22}
J.~A. Backer, D.~Eggink, S.~P. Andeweg, I.~K. Veldhuijzen, N.~van Maarseveen,
  K.~Vermaas, B.~Vlaemynck, R.~Schepers, S.~van~den Hof, C.~B. Reusken and
  J.~Wallinga,
\newblock \emph{Shorter serial intervals in SARS-CoV-2 cases with Omicron ba.1
  variant compared with delta variant, the netherlands, 13 to 26 december
  2021.},
\newblock Euro Surveill. \textbf{27(6)}, 2200042 (2022),
\newblock \doi{10.2807/1560-7917.ES.2022.27.6.2200042}.

\bibitem{CD15}
D.~Champredon and J.~Dushoff,
\newblock \emph{Intrinsic and realized generation intervals in
  infectious-disease transmission.},
\newblock Proc Biol Sci. \textbf{282}, 1821 (2015),
\newblock \doi{10.1098/rspb.2015.2026}.

\bibitem{Alietal20}
S.~T. Ali, L.~Wang, E.~H.~Y. Lau, X.-K. Xu, Z.~Du, Y.~Wu, G.~M. Leung and B.~J.
  Cowling,
\newblock \emph{Serial interval of SARS-CoV-2 was shortened over time by
  nonpharmaceutical interventions},
\newblock Science \textbf{369}(6507), 1106 (2020),
\newblock \doi{10.1126/science.abc9004},
\newblock \eprint{https://www.science.org/doi/pdf/10.1126/science.abc9004}.

\bibitem{Parketal21}
S.~W. Park, K.~Sun, D.~Champredon, M.~Li, B.~M. Bolker, D.~J.~D. Earn, J.~S.
  Weitz, B.~T. Grenfell and J.~Dushoff,
\newblock \emph{Forward-looking serial intervals correctly link epidemic growth
  to reproduction numbers},
\newblock Proceedings of the National Academy of Sciences \textbf{118}(2)
  (2021),
\newblock \doi{10.1073/pnas.2011548118},
\newblock \eprint{https://www.pnas.org/content/118/2/e2011548118.full.pdf}.

\bibitem{Manietal22}
M.~Manica, A.~De~Bellis, G.~Guzzetta, P.~Mancuso, M.~Vicentini, F.~Venturelli,
  E.~Bisaccia, M.~Litvinova, P.~Poletti, V.~Marziano, A.~Zardini, V.~d'Andrea
  \emph{et~al.},
\newblock \emph{Intrinsic generation time of the SARS-CoV-2 omicron variant: An
  observational study of household transmission.},
\newblock Preprint with the Lancet \textbf{submitted} (2022),
\newblock \doi{10.2139/ssrn.4068368}.

\bibitem{KMKW27}
W.~O. Kermack, A.~G. McKendrick and G.~T. Walker,
\newblock \emph{A contribution to the mathematical theory of epidemics},
\newblock Proceedings of the Royal Society of London. Series A, Containing
  Papers of a Mathematical and Physical Character \textbf{115}(772), 700
  (1927),
\newblock \doi{10.1098/rspa.1927.0118},
\newblock
  \eprint{https://royalsocietypublishing.org/doi/pdf/10.1098/rspa.1927.0118}.

\bibitem{GrFr08}
N.~C. Grassly and C.~Fraser,
\newblock \emph{Mathematical models of infectious disease transmission},
\newblock Nature Reviews Microbiology \textbf{6}, 477 (2008),
\newblock \doi{10.1038/nrmicro1845}.

\bibitem{Braetal21}
L.~T. Brandal, E.~MacDonald, L.~Veneti, T.~Ravlo, H.~Lange, U.~Naseer,
  S.~Feruglio, K.~Bragstad, O.~Hungnes, L.~E. Ødeskaug, F.~Hagen, K.~E.
  Hanch-Hansen \emph{et~al.},
\newblock \emph{Outbreak caused by the SARS-CoV-2 omicron variant in norway,
  november to december 2021.},
\newblock Euro Surveill. \textbf{26(50)}, 2101147 (2021),
\newblock \doi{10.2807/1560-7917.ES.2021.26.50.2101147}.

\bibitem{Songetal22}
J.~Song, J.~Lee, K.~M and et~al.,
\newblock \emph{Serial intervals and household transmission of SARS-CoV-2
  omicron variant, south korea, 2021.},
\newblock Emerg Infect Dis. \textbf{28(3)}, 756 (2022),
\newblock \doi{10.3201/eid2803.212607}.

\bibitem{aHB22}
M.~an~der Heiden and U.~Buchholz,
\newblock \emph{Serial interval in households infected with SARS-CoV-2 variant
  b.1.1.529 (Omicron) are even shorter compared to delta},
\newblock Epidem. and Infect. \textbf{To appear} (2022).

\bibitem{IPN22}
K.~Ito, C.~Piantham and H.~Nishiura,
\newblock \emph{Estimating relative generation times and relative reproduction
  numbers of Omicron ba.1 and ba.2 with respect to delta in denmark},
\newblock medRxiv  (2022),
\newblock \doi{10.1101/2022.03.02.22271767},
\newblock
  \eprint{https://www.medrxiv.org/content/early/2022/03/04/2022.03.02.22271767.full.pdf}.

\bibitem{Fra07}
C.~Fraser,
\newblock \emph{Estimating individual and household reproduction numbers in an
  emerging epidemic},
\newblock PLOS ONE \textbf{2}(8), 1 (2007),
\newblock \doi{10.1371/journal.pone.0000758}.

\bibitem{BLGdA11}
M.~Bottiglieri, E.~Lippiello, C.~Godano and L.~de~Arcangelis,
\newblock \emph{Comparison of branching models for seismicity and likelihood
  maximization through simulated annealing},
\newblock Journal of Geophysical Research: Solid Earth \textbf{116}(B2), n/a
  (2011),
\newblock \doi{10.1029/2009JB007060},
\newblock B02303.

\bibitem{LGdAMG14}
E.~Lippiello, F.~Giacco, L.~de~Arcangelis, W.~Marzocchi and C.~Godano,
\newblock \emph{Parameter estimation in the {ETAS} model: Approximations and
  novel methods},
\newblock Bulletin of the Seismological Society of America \textbf{104}(2), 985
  (2014),
\newblock \doi{10.1785/0120130148},
\newblock \eprint{http://www.bssaonline.org/content/104/2/985.full.pdf+html}.

\end{thebibliography}

\end{document}